\documentclass[prb,amsmath,amssymb,twocolumn,floatfix,citeautoscript]{revtex4}
\usepackage{bm}
\usepackage{epsfig}

\begin{document}

\title{RKKY Interaction and Intrinsic Frustration in Non-Fermi Liquid Metals}

\author{Jian-Huang She, and A. R. Bishop}

\affiliation{Theoretical Division, Los Alamos National Laboratory, Los Alamos, NM, 87545, USA.}

\begin{abstract}
We study RKKY interaction in non-Fermi liquid metals. We find that the RKKY interaction mediated by non-Fermi liquid metals can be of much longer range than for a Fermi liquid. The oscillatory nature of RKKY interaction thus becomes more important in non-Fermi liquids, and gives rise to enhanced frustration when the spins form a lattice. Frustration suppresses the magnetic ordering temperature of the lattice spin system. Furthermore, we find that the spin system with longer range RKKY interaction can be described by the Brazovskii model, where the ordering wavevector lies on a shell with constant radius. Strong fluctuations in such model lead to a first-order phase transition and/or glassy phase. This may explain some recent experiments where glassy behavior was observed in stoichiometric heavy fermion material close to a ferromagnetic quantum critical point.
\end{abstract}

\date{\today \ [file: \jobname]}

 \maketitle

\textbf{\emph{ Introduction:}}  When magnetic moments are placed in a metal, the conduction electrons mediate an indirect interaction between these moments. Such a long rang interaction is called the Ruderman-Kittel-Kasuya-Yosida (RKKY) interaction. RKKY interaction plays crucial roles in, e.g., heavy fermions, diluted magnetic semiconductors, graphene. The usual derivation of the RKKY interaction is based on the assumption that the conduction electrons form a Landau Fermi liquid. However many strongly correlated electron systems show non-Fermi liquid behavior, e.g. cuprates, heavy fermions, pnictides. The question we ask here is what is the form of RKKY interaction in a non-Fermi liquid metal, and what are the consequencies.

Of particular interest are heavy fermion systems, where local moements couple to the conduction electrons. The Doniach phase diagram with competing Kondo coupling and RKKY interaction has been the paradigm for heavy fermions for decades [\onlinecite{Doniach77}]. In the last few years, as experimental results accumulate, there is a growing necessity to go beyond the Doniach phase diagram. Frustration or the quantum zero point energy has been proposed as a new dimension in the global phase diagram of heavy fermions [\onlinecite{Senthil03, Senthil04, Si06, Coleman10, Custers12}]. One obvious origin of frustration is frustration of lattice structure itself. However such geometric frustration is not universally observed in heavy fermion materials. Here we propose that the non-Fermi liquid nature of conduction electrons in the Kondo liquid phase leads to intrinsic frustration for the localized spin degrees of freedom. This provides a more universal source of frustration.

\begin{figure}
\begin{centering}
\includegraphics[width=0.43\linewidth]{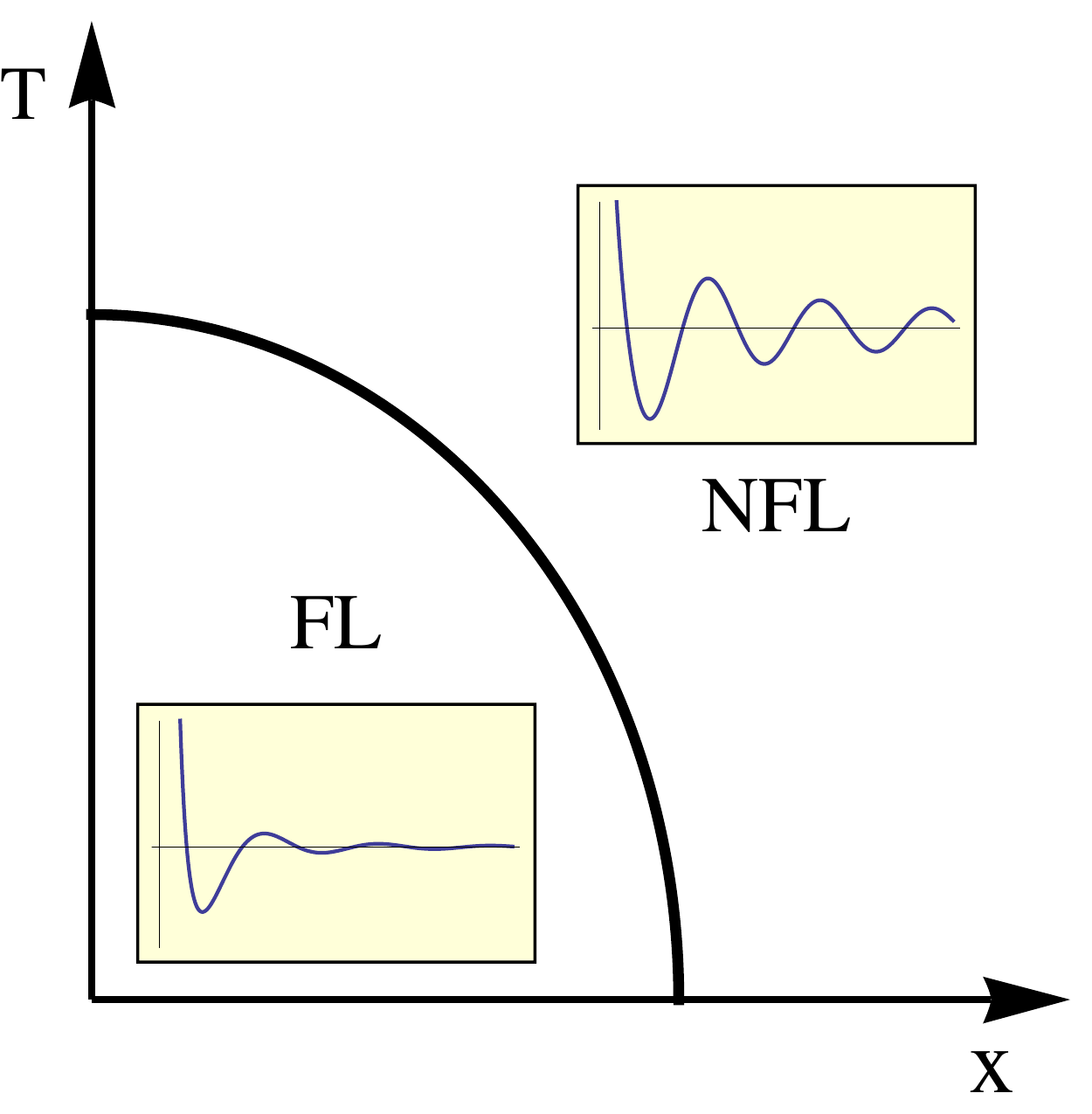} 
\includegraphics[width=0.43\linewidth]{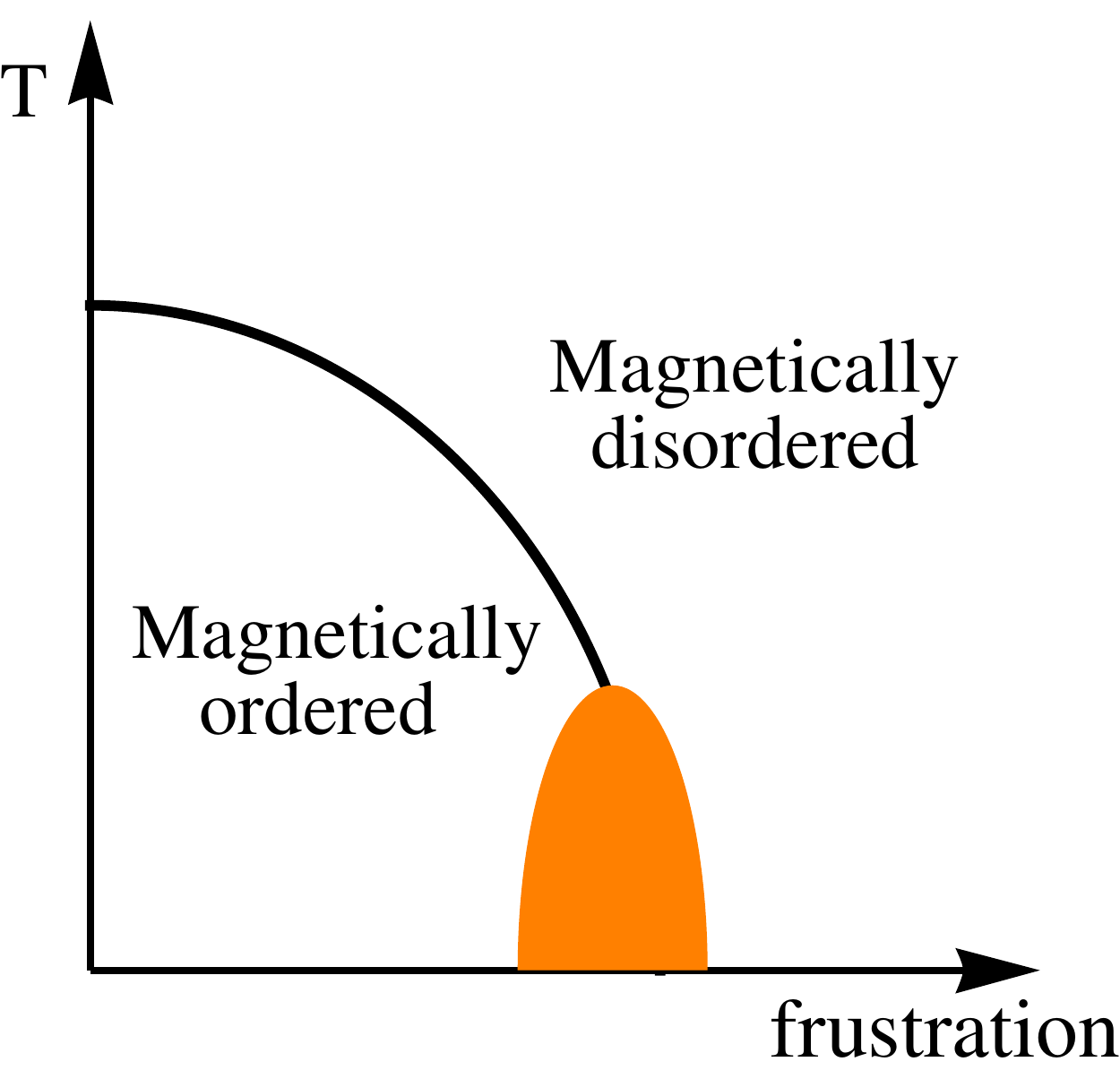} 
\end{centering}
\caption{Schematic electronic and magnetic phase diagrams. Distance dependence of the RKKY interaction is shown in the insets. In the non-Fermi liquid region, RKKY interaction is of longer range, leading to frustration. The magnetic transition temperature decreases with increasing frustration, and new phases can emerge near the QCP.}
\label{Phase}
\end{figure}

Our approach is based on the idea of quantum criticality and non-Fermi liquid (NFL) behavior. The standard picture is that the critical fluctuations near a quantum critical point (QCP) leads to NFL behavior. Here we depart from this picture by starting with the assumption that in a certain range of the parameter space, the itinerant electrons form a NFL state. We then proceed to study its consequences on other degrees of freedom, e.g. the localized spins. Focusing on the regime with small Kondo coupling, i.e. a small Fermi surface, we find that the magnetic transition temperature will be reduced by the frustration resulting from longer-range RKKY interaction produced by NFL itinerant electrons. Furthermore, we find that the putative ferromagnetic (FM) QCP may be replaced by a first-order phase transition or a glassy phase [\onlinecite{Mydosh, Hertz}] (see Fig.~\ref{Phase}).

\textbf{\emph{ Formalism:}} We start with the Kondo lattice model, $H=H_C+H_K$. Here $H_C$ is the conduction electron Hamiltonian, and ususally only the hopping term is included. The Kondo coupling between conduction electrons and localized spins is of the form, $H_K=-\frac{J_K}{2}\sum_{i\alpha\beta} {\bm S}_i\cdot c^{\dagger}_{i\alpha}{\boldsymbol \sigma}_{\alpha\beta}c_{i\beta}$. We depart from the usual approach by considering the conduction electrons to be strongly interacting themselves, i.e. $H_C=H_C^{(0)}+H_C^{({\rm int})}$. One way to motivate this is to consider the phenomenological two fluid model [\onlinecite{Fisk04, Pines04, Pines0802, Pines08}]. In many heavy fermion systems, below the coherence temperature $T^*$, the experimental results can be understood in terms of the two fluid model, with one component the itinerant heavy electrons, and the other component local moments. The heavy electron Kondo liquid is not a simple Fermi liquid, e.g. the specific heat is logarithmically enhanced at low temperature. One has a model of interacting itinerant electrons coupled with localized spins.

The itinerant electrons induce a RKKY type interaction among the localized moments:
\begin{equation}
H_{\rm RKKY}=\sum_{ij}J^{ab}_{ij}  S^a_i  S^b_j.
\label{spin-H}
\end{equation}
Here the coupling $J^{ab}_{ij} =-\frac{J_K^2}{4}\chi^{ab}_{ij}$ [\onlinecite{Ruderman54, Loss07, Loss08}], is determined by the static spin susceptibility of the conduction electrons
 \begin{equation}
\chi^{ab}_{ij} =-\frac{i}{\hbar}\int_0^{\infty}\langle [s^a({\bm r}_i,t), s^b({\bm r}_j,0) ]\rangle e^{-\eta t}  dt,
\end{equation}
with the electron spin $s^a({\bm r}_i)=\sum_{\alpha\beta}c^{\dagger}_{i\alpha} \sigma^a_{\alpha\beta}c_{i\beta}$ and $\eta=0^+$. If the conduction electrons are in the paramagnetic state, the spin susceptibility is isotropic and diagonal, i.e. $\chi^{ab}_{ij}=\chi({\bm r}_{ij})\delta^{ab}$. For Fermi liquids, the spin susceptibility behaves as $\chi({\bm r})\sim (1/r^d)\cos(2k_Fr+\theta_0)$ at long distances, with $d$ the spatial dimension. This leads directly to the standard form of the RKKY interaction. The exponent $d$ results from the sharp jump in the momentum distribution $n({\bm k})$, characteristic of Fermi liquids. For non-Fermi liquid metals, the RKKY interaction can have qualitatively different behavior. We still assume the existence of a Fermi surface, i.e. singularity in $n({\bm k})$, thus the spin susceptibility still has $2k_F$ oscillation. The exponent can take a different value. Thus we have $\chi({\bm r})\sim (1/r^\alpha)\cos(2k_Fr+\theta_0)$. More detailed studies of the NFL spin susceptibility will be presented below.

Consider placing a lattice of spins in the non-Fermi liquid metal. We focus on the effect of the RKKY interaction to the spin system, and will not consider the competition between Kondo coupling and RKKY interaction [\onlinecite{Doniach77}]. This can be achieved by assuming the spins to be classical, or considering only the part of the phase diagram with a small Fermi surface. With ${\bm S}({\bm q})=(1/N)\sum_i{\bm S}_i e^{i{\bm q}\cdot {\bm r}_i}$, one has in momentum space, $H=\sum_{\bm q}F({\bm q}){\bm S}({\bm q})\cdot {\bm S}(-{\bm q})$, where 
\begin{equation}
F({\bm q})=\frac{1}{N}\sum_{{\bm r}_i\neq 0}J({\bm r}_i)e^{i{\bm q}\cdot {\bm r}_i},
\label{FqSum}
\end{equation}
with ${\bm r}_i$ defined on the lattice. The ordering wavevector in the ground state is determined by minimizing the function $F({\bm q})$.

 For the conventional three dimensional RKKY interaction mediated by a Fermi liquid, this problem has been studied in [\onlinecite{Mattis62}], where different phases have been identified as the conduction electron density changes. At small $k_Fa$, where $a$ is the lattice constant, the ground state is ferromagnetic. As $k_Fa$ increases, antiferromagnetic phases with different ordering wavevectors appear. In the case $k_Fa\to 0$, the above summation can be replaced by an integral, and $F({\bm q})\sim -\chi(q)$. The ordering wavevector is thus determined by maximizing the static spin susceptibility.

 Now we consider in more detail what is the form of the static spin susceptibility in a NFL metal. When vertex corrections can be ignored, the spin susceptibility can be calculated from the fermion bubble, with $\chi_{ab}(q)\sim \int \sigma^aG(k+q)\sigma^bG(k)$. When the momentum distribution $n({\bm k})$ has a weaker singularity than a jump at $k_F$, e.g. a kink, the Friedel oscillation decays faster than that of Fermi liquids (see Appendix). Then one expects $\chi(r)$ and $J(r)$ to decay faster than that of Fermi liquids. An interesting question is whether it is possible to have longer range RKKY interactions, which would generate the desired frustration among the spins [\onlinecite{Senthil03, Senthil04, Si06, Coleman10, Custers12}]. We will present two models of NFL metals that can give rise to such behavior.

\textbf{\emph{ Longer range RKKY interaction in 1-d:}} First, as a proof of principle that RKKY interaction in a strongly interacting electron system can be of longer range than in a free system, let us first consider one dimension. In 1-d, RKKY interaction mediated by free electrons is of the form $J(r)\sim {\rm Si}(2k_F r)-\frac{\pi}{2}$, with the sine integral function ${\rm Si}(x)$. At large distance one has $J(r)\sim \cos(2k_Fr)/r$. In momentum space, one has $\chi(q)\sim (1/q)\ln|\frac{q+2k_F}{q-2k_F}|$, with a maximum at $q=2k_F$.

 The low energy dynamics of interacting electrons in 1-d is described by the Luttinger liquid theory. Due to spin-charge separation, the conduction electron Hamiltonain can be written as a summation of the two channels [\onlinecite{Gogolin}],
\begin{equation}
H_C=\sum_{\alpha=c,s} \frac{v_{\alpha}}{2}\int dx[g_{\alpha}\Pi_{\alpha}^2+g_{\alpha}^{-1}(\partial_x\theta_{\alpha})^2],
\end{equation}
with $v_c$ and $v_s$ the velocity of charge and spin density wave respectively. The charge interaction constant $g_c=1$ for noninteracting fermions, $g_c< 1$ for repulsive interaction, and $g_c>1$ for attractive interaction. We are interested in the case with repulsive interaction. The spin interaction constant $g_s=1$ in the presence of SU(2) spin symmetry. The oscillating part of the spin correlation function is [\onlinecite{Gogolin}]
\begin{equation}
\langle{\bm s}(x,\tau)\cdot{\bm s}(0,0)\rangle\sim \frac{\cos(2k_F x)}{|\tau+ix/v_c|^{g_c}|\tau+ix/v_s|^{g_s}}.
\end{equation}
The RKKY interaction, determined from the static spin susceptibility, is of the form
\begin{equation}
J(x)\sim \int d\tau\frac{\cos(2k_F x)}{|\tau+ix/v_c|^{g_c}|\tau+ix/v_s|^{g_s}}\sim\frac{\cos(2k_F x)}{x^{g_c+g_s-1}}.
\end{equation}
For $g_c<1$, $g_s=1$, the exponent $\alpha=g_c+g_s-1<d=1$. The RKKY interaction mediated by a Luttinger liquid is thus of longer range than that mediated by a non-interacting Fermi gas \footnote{RKKY interaction mediated by 1-d Luttinger liquid has been studied in [\onlinecite{Egger96}], where the correlation function of the conduction electron spin and the impurity spin, $\langle s_z(x)S_z(0)\rangle$ was considered.}.

\begin{figure}
\begin{centering}
\includegraphics[width=0.45\linewidth]{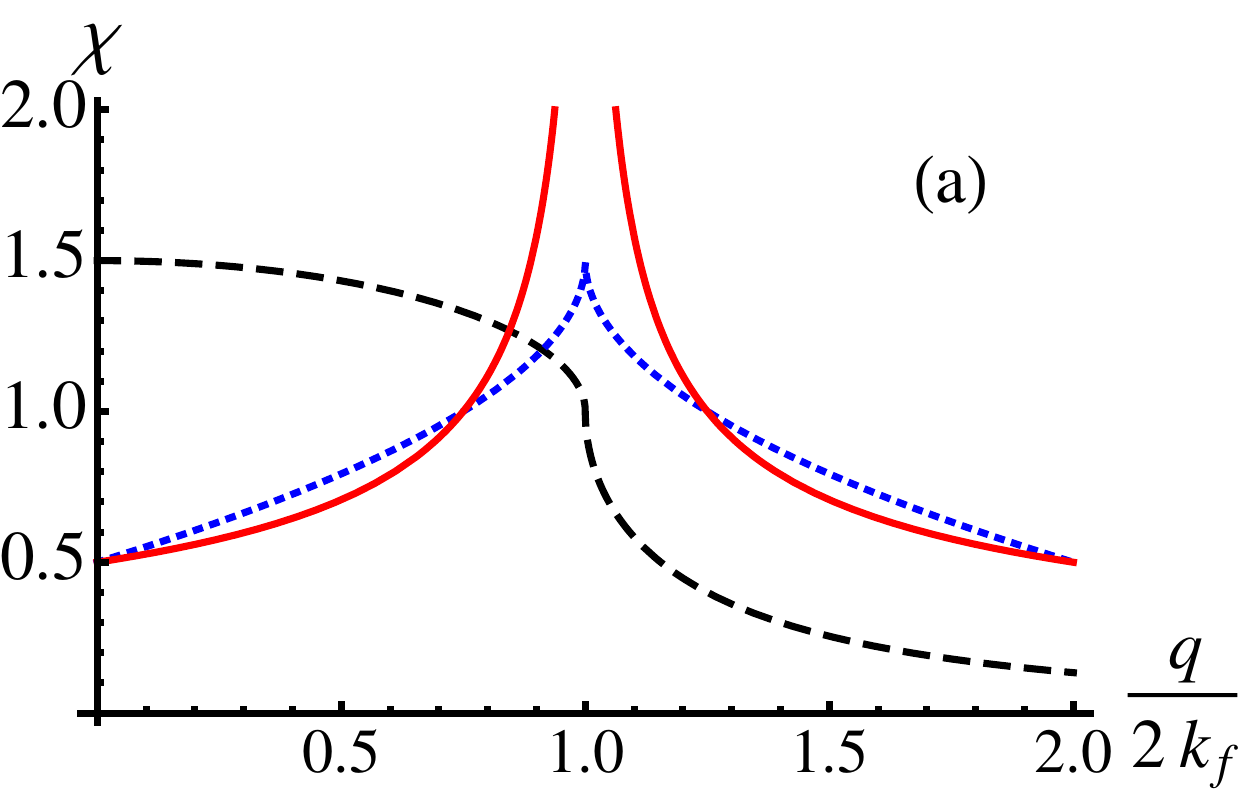} 
\includegraphics[width=0.45\linewidth]{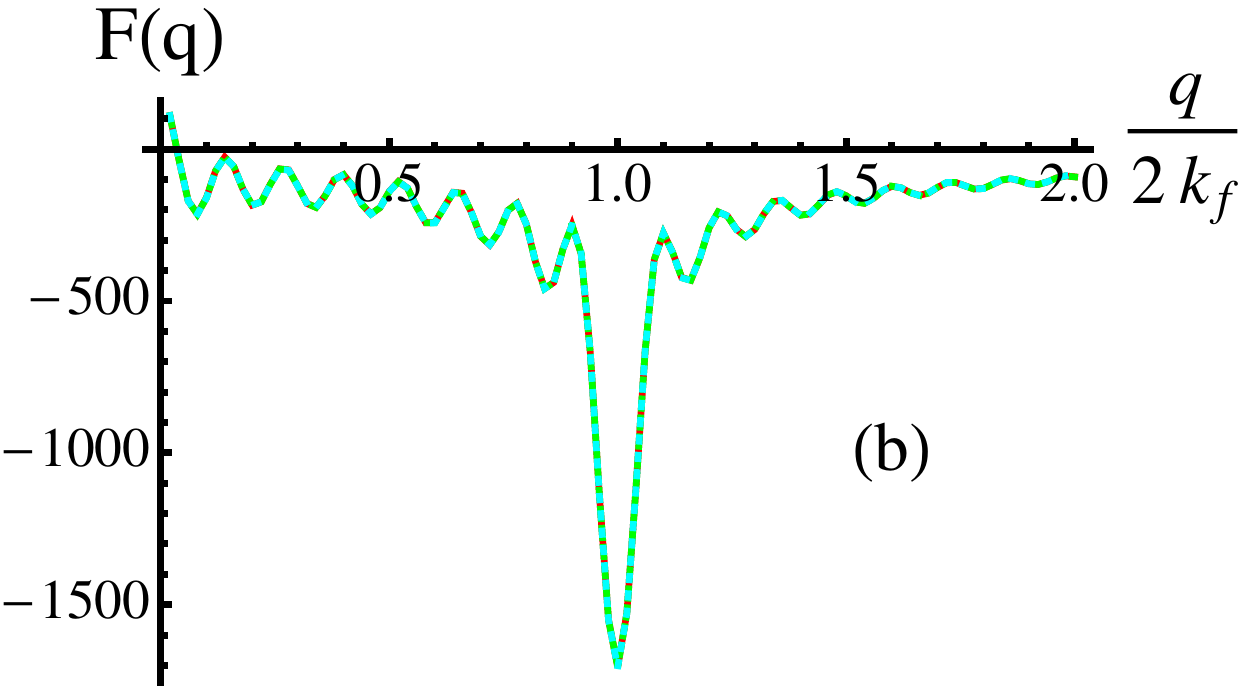} 
\end{centering}
\caption{(a) The static spin susceptibility $\chi(q)$ as function of momentum for Fermi liquid (dashed, black) and the gauge-fermion model with $\sigma<1/3$ (dotted, blue) and $\sigma>1/3$ (solid, red). (b) $F(q)$ as function of momentum for angles $\theta=0, \pi/6, \pi/4$. The curves for different angles are almost identical. Here the spins form a square lattice, and $\sigma=1/2, k_Fa=0.2$.}
\label{chi}
\end{figure}

\textbf{\emph{ Spin susceptibility in 2-d:}} Now we consider two dimenional metals. For free electrons, the static spin susceptibility reads
\begin{equation}
\chi(q)=\left\{ \begin{array}{lcl} \chi_0& \mbox{for} & q<2k_F  \\   \chi_0\left[ 1-\sqrt{1-(2k_F/q)^2} \right]& \mbox{for} &  q>2k_F,
\end{array}\right.
\label{2d-FF}
\end{equation}
with $\chi_0=1/\pi$, which has a one-sided square-root singularity. The RKKY interaction is thus of the form $J(r)\sim \sin(2k_Fr)/r^2$. \footnote{For graphene with a Dirac spectrum, the RKKY interaction decays faster than that of the free electron gas, with $J(r)\sim [1+\cos(2{\bm K}_D\cdot{\bm r})]/r^3$, where ${\bm K}_D$ is the reciprocal vector for the Dirac points. This is consistent with the classification in Ref.[\onlinecite{She09}] of free fermions as having marginal susceptibility, Dirac materials having irrelevant susceptibility, and quantum critical metals having relevant susceptibility.} For 2-d Fermi liquid, including higher order diagrams, there is also a square-root singularity for $q<2k_F$, with $\chi(q)=\chi(2k_F)+\chi^{\rm sing}(q)$ [\onlinecite{Chubukov9302}], and
\begin{equation}
\chi^{\rm sing}(q)={\cal A}\sqrt{1-(q/2k_F)^2}.
\label{2d-FL}
\end{equation}
The new singularity gives contribution $\delta\chi(r)\sim [(2k_Fr)\cos(2k_Fr)-\sin(2k_Fr)]/r^3$, and will not change the long distance behavior of the RKKY interaction.

 A prototype of non-Fermi liquid metal in higher dimensions is the system of 2-d degenerate fermions interacting via a singular gauge interaction [\onlinecite{Polchinski94, Nayak94, Nayak9402, Kim94, Altshuler94, Lee09, Sachdev10, Senthil10}], where the presence of the gauge interaction leads to singular $2k_F$ response [\onlinecite{Altshuler94}]. The fermion $2k_F$ vertex $\Gamma_{2k_F}$ has a power law dependence on frequency, with $\Gamma_{2k_F}\sim \left(\frac{E_F}{\omega} \right)^{\sigma}\Gamma^0_{2k_F}$. The exponent is of the form $\sigma= \frac{1}{2N}+\frac{1}{2\pi^2N^2}\ln^3N+{\cal O}\left(\frac{1}{N^2}\right)$ for large $N$, and $\sigma=\frac{16\sqrt{2}}{9\pi\sqrt{N}}+{\cal O}\left(1\right)$ for small $N$. Here the spin index is generalized to take values from $1$ to $N$. Taking $N=2$, one obtains $\sigma=0.35$ from the large-N expansion, and $\sigma=0.56$ in the small-N limit.

 The spin susceptibility is calculated from the polarization bubble with vortex corrections, $\chi({\bm q},\omega)\simeq \Pi({\bm q},\omega)=\int d{\bm p}d\epsilon G({\bm p}+{\bm q}/2,\epsilon+\omega/2)G({\bm p}-{\bm q}/2,\epsilon-\omega/2)\left[\Gamma_{\epsilon{\bm p}}({\bm q},\omega) \right]^2$. For $\sigma<1/3$, the static spin susceptibility is of the form [\onlinecite{Altshuler94}]
  \begin{equation}
\chi(q)\sim \chi_0 -{\cal C}|q-2k_F|^{1-3\sigma},
\label{2d-NFL1}
\end{equation}
and for $\sigma>1/3$ one has [\onlinecite{Altshuler94}]
   \begin{equation}
\chi(q)\sim \frac{1}{|q-2k_F|^{3\sigma-1} },
\label{2d-NFL2}
\end{equation}
with a singularity at $q=2k_F$ (see Fig.~\ref{chi}(a)). Fourier transforming to real space, we find
\begin{equation}
\chi({\bm r})\sim \int \frac{1}{|q-2k_F|^{3\sigma-1}}J_0(qr)qdq\sim \frac{\cos(2k_Fr-\theta_0)}{r^{5/2-3\sigma}}.
\end{equation}
 The exponent $\alpha=5/2-3\sigma$ can be much smaller than the space dimension $d=2$.

 More generally, for non-Fermi liquid metals, one can employ a scaling theory for the susceptibility (see e.g. [\onlinecite{Senthil08, She09}]). Assuming the existence of a Fermi surface, the static spin susceptibility generally has a power law behavior near $q=2k_F$, with $\chi({\bm q})\sim |q-2k_F|^\nu$. For $\nu<1/2$, one has a stronger singularity than the Fermi liquid case, and the RKKY interaction is of longer range.

\textbf{\emph{Longer range RKKY interaction in 2-d:}} Let us now consider the ground state of the spins embedded in the 2-d metals with small $k_Fa$. For the Fermi liquid case (Eqs.(\ref{2d-FF},\ref{2d-FL})), $\chi(q)$ increases monotonically with decreasing $q$ (see Fig.~\ref{chi}). The ground state is ferromagnetic. For non-Fermi liquid (Eqs.(\ref{2d-NFL1}),(\ref{2d-NFL2})), the maximum of $\chi(q)$ is at $q=2k_F$, indicating an instability of the ferromagnetic state. More precisely, one can calculate the interaction $F(q)$ by first Fourier transforming $\chi({\bm q})$ to real space to get $\chi({\bm r})$, and then performing the lattice summation in Eq.~(\ref{FqSum}). The result for $\sigma=1/2$ is shown in Fig.~\ref{chi}(b). One can see that $F(q)$ has a minimum at $q=2k_F$. The singularity in $\chi(q)$ is smeared out by the lattice effect. 

Another observation is that $F({\bm q})$ has a very weak dependence on the direction of momentum. In Fig.~\ref{chi}(b), $F({\bm q})$ for the three different angles are almost indistinguishable. With the minimum of $F({\bm q})$ at $q_0=2k_F$, the ordering wavevector of the lattice spin system lies on a shell of radius $2k_F$. Expanding $F({\bm q})$ around $q_0$, one obtains the Brazovskii model [\onlinecite{Brazovskii75}],
\begin{equation}
H=\sum_{\bm q}\left[b_0 +D\left(|{\bm q}|-q_0\right)^2\right]{\bm S}({\bm q})\cdot {\bm S}(-{\bm q}).
\end{equation}
Brazovskii found that the large phase space available for fluctuations around a shell of minima leads to a first-order phase transition [\onlinecite{Brazovskii75}]. It has been found experimentally that putative FM-QCPs are replaced by first order transitions at low temperatures in several transition metal compounds, e.g. MnSi, ZrZn$_2$, and heavy fermion systems, e.g. UGe$_2$, UCoAl, UCoGe (see [\onlinecite{Pfleiderer09}] and references therein). It was realized earlier that competing orders [\onlinecite{She10}] as well as fluctuations [\onlinecite{Chubukov04, Green09, Green1201, Green1202}] can lead to first order quantum phase transitions. Here we find a new mechanism where the frustration resulting from NFL behavior generates first order transitions.

A further observation is that the extensive configurational entropy in the Brazovskii model should  lead to slow dynamics and glassiness [\onlinecite{Schmalian00, Schmalian01, Schmalian03, Nussinov09}]. \footnote{These two pictures are not necessarily in conflict with each other. The identification of first-order transition is about static properties, while the glass transition is ultimately about dynamics. So the two pictures are concerned with different time scales. Furthermore, glassy dynamics is usually observed near a first order phase transition when the system is supercooled to the low temperature ordered phase.} Glassy correlations emerge when the correlation length $\xi=(D/b)^{1/2}$ becomes of order the modulation length $l_0=2\pi/q_0$ [\onlinecite{Schmalian01}]. The parameter $b$ needs to be determined self-consistently. Within the large-N approximation, and including a small quartic term with coupling $u$, we have
\begin{equation}
b=b_0+uT\int\frac{d^2{\bm q}}{(2\pi)^2}{\cal G}(q),
\end{equation}
with the Green's function ${\cal G}({\bm q})=1/[b+D\left(q-q_0\right)^2]$. The condition $\xi/ l_0\sim 1$ then determines the temperature where glassy behavior sets in to be $T_g\simeq \frac{2\pi D^2}{u}\frac{q_0^2-b_0/D}{c_1-\log(q_0a)}$, with the coefficient $c_1$ of order unity and momentum cutoff $\Lambda\sim a^{-1}$. We notice that here $T_g$ depends logarithmically on cutoff instead of the $1/\Lambda$ dependence for the 3-d model considered in [\onlinecite{Schmalian01}].

Glassy spin dynamics was recently observed in the heavy fermion system CeFePO [\onlinecite{Lausberg12}]. CeFePO is a layered Kondo lattice system, in close proximity to a FM QCP. Spin-glass-like freezing was detected in the ac susceptibility, specific heat and muon-spin relaxation [\onlinecite{Lausberg12}]. The glass behavior in such a stoichiometric system points to new mechanisms that do not reply on external randomness. Our model provides such a possibility (see [\onlinecite{Parisi9401, Parisi9402, Parisi95, Hertz95, Chandra96, Loh04}] and references therein for earlier attempts to obtain glass behavior from frustrated deterministic models).

\textbf{\emph{Away from QCP:}} Having identified a glass transition near the QCP, we proceed to study the behavior of the lattice spin system away from QCP using a random exchange model that is widely used to describe spin glasses. Due to the cosin function, the RKKY interaction changes sign and magnitude with distance. It can be well approximated by a random interaction [\onlinecite{Abrikosov78, Gulacsi97, Gulacsi98, Gulacsi05}], $J_{ij}\sim \frac{J_K^2}{4}\frac{\epsilon_{ij}}{r^{\alpha}}$, where $\epsilon_{ij}$ is a random variable with cosine distribution $P(\epsilon_{ij})=(1/\pi ) (1-\epsilon_{ij}^2)^{-1/2}$. 
 
When the itinerant electrons are away from the QCP, there is a crossover to the Fermi liquid behavior at low energy, or equivalently long distance, where the RKKY interaction is substantially reduced. We will assume for simplicity that the RKKY interaction can be neglected beyond a crossover scale $r_{\rm FL}$. Then the exchange interaction is of the form 
\begin{equation}
\label{random}
J_{ij}=\left\{ \begin{array}{ccl}  {\cal A}\epsilon_{ij}/|{\mathbf r}_i-{\mathbf r}_j|^{\alpha}& \mbox{for} & |{\mathbf r}_i-{\mathbf r}_j|<r_{\rm FL}  \\0& \mbox{for} &  |{\mathbf r}_i-{\mathbf r}_j|>r_{\rm FL}.
\end{array}\right.
\end{equation}
When $r_{\rm FL}$ becomes of order the lattice constant, only the nearest neighbor interactions survive, i.e. $H=J\sum_{<ij>}{\mathbf S}_i \cdot {\mathbf S}_j$, and the spins are magnetically ordered. As $r_{\rm FL}$ increases, the ordering temperature will be reduced by frustration.

 A simpler model that illustrates essentially the same effect of suppression of ordering by frustration is the Sherrington-Kirkpatrick model [\onlinecite{Kirkpatrick75, Sherrington75}].  Consider here ferromagnetic ordering. We start with a mean field type Hamiltonian $H=-J_0\sum_{(ij)}{\bm S}_i \cdot {\bm S}_j$, with $J_0>0$, and each spin interacts with $z$ neighouring spins. The spins order ferromagnetically below the transition temperature $T_c^{(0)}={\tilde J}_0S(S+1)/6$, with ${\tilde J}_0=zJ_0$. This correpondes to the case far away from the QCP.

Then we add to the above mean field ferromagnetic model random exchange interactions to model the frustration effect when approaching a QCP.  The new Hamiltonian can be written as $H=-\sum_{(ij)}J_{ij}{\bm S}_i \cdot {\bm S}_j$, where the interaction $J_{ij}$ is distributed according to $P(J_{ij})=\frac{1}{\sqrt{2\pi J^2}}\exp\left[ -\frac{(J_{ij}-J_0)^2}{2J^2} \right]$ [\onlinecite{Kirkpatrick75, Sherrington75}]. This model is readily solved by the replica technique [\onlinecite{Sherrington75, Gabay81}], and the transition temperature to ferromagnetism is reduced by the random interactions, with the result [\onlinecite{Sherrington75, Montgomery70}]
\begin{equation}
T_c=T_c^{(0)}\left[ \frac{1}{2}+\frac{1}{2}\sqrt{1-\frac{3}{S(S+1)}\frac{{\tilde J}^2}{{\tilde J}_0^2}} \right],
\end{equation}
where we have defined ${\tilde J}=z^{1/2}J$.

\begin{figure}
\begin{centering}
\includegraphics[width=0.6\linewidth]{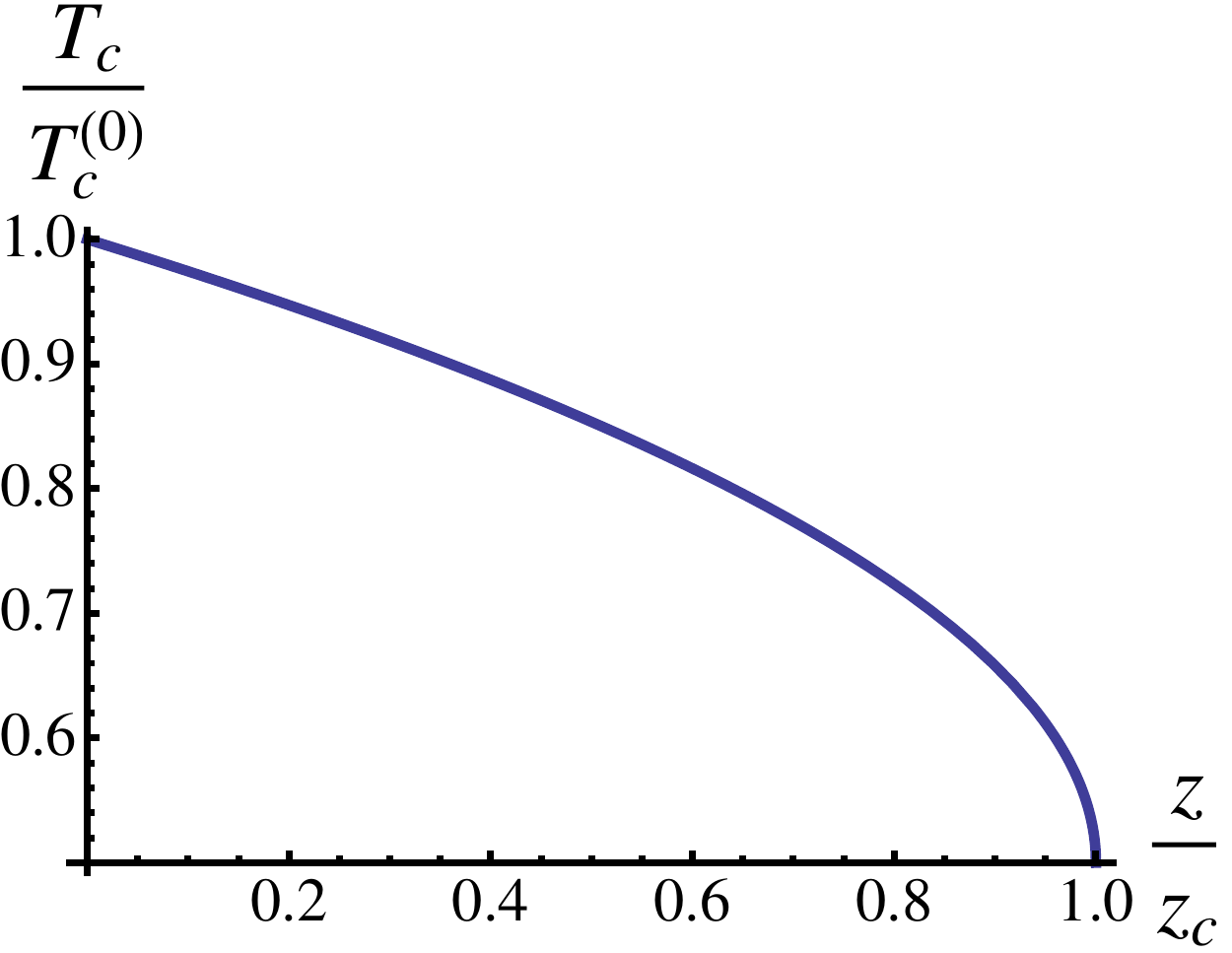} 
\end{centering}
\caption{Ferromagmetic transition temperature as function of range of random exchange interaction.}
\label{FM}
\end{figure}

We fix the mean field ordering temperature in the absence of random exchange interaction $T_c^{(0)}$ and the variance of the random distribution $J$, so that $z$ is a measure of the range of random exchange interaction, i.e. $r_{\rm FL}$ in Eq.(\ref{random}).  We can define $z_c=(S(S+1)/3){\tilde J}_0^2/J^2$, and write $T_c$ in the form
\begin{equation}
T_c=T_c^{(0)}\left[ \frac{1}{2}+\frac{1}{2}\sqrt{1-\frac{z}{z_c}} \right],
\end{equation}
which is plotted in Fig.~\ref{FM}. One can see that with increasing range of random exchange interaction, the FM ordering temperature decreases. This then translates to the picture that when approaching the QCP, as RKKY interaction becomes of longer range, magnetic ordering is suppressed (see Fig.~\ref{Phase}).

\textbf{\emph{ Conclusions:}} We have studied the RKKY interaction in non-Fermi liquid metals. The basic picture we find is summarized in Fig.~\ref{Phase}. In the non-Fermi liquid phase, when including vertex corrections, the RKKY interaction can be of longer range than in a Fermi liquid. Longer range RKKY interaction leads to frustration for the lattice spin system placed in such a NFL metal. Magnetic ordering will be suppressed by frustration, and novel behavior may emerge near the putative QCP. In particular, the continuous second-order phase transitions may be replaced by first-order transtions. Glassy dynamics may occur near the QCP without invoking disorder. One candidate material for such glass behavior is the heavy fermion system CeFePO.
We focused here on FM QCP. One can also generalize the whole procedure to AFM QCP by increasing $k_Fa$. Another interesting question is the competition between the Kondo coupling and the longer range RKKY interaction.

We acknowledge useful discussions with Sasha Balatsky, Cristian D. Batista, Andrey Chubukov, Matthias Graf, Jason T. Haraldsen, John Hertz, John Mydosh, Stephen Powell, Jan Zaanen, and Jian-Xin Zhu. This work was supported by the U.S. Department of Energy under contract DE-AC52-06NA25396 at Los Alamos National Laboratory through the Basic Energy Sciences program, Materials Sciences and Engineering Division. 

\section*{Appendix: Friedel oscillation}

We consider here the Friedel oscillation for a non-Fermi liquid metal. When the momentum distribution $n({\bm k})$ has a weaker singularity than a jump at $k_F$, e.g. a kink, the Friedel oscillation decays faster than that of Fermi liquids. Consider for simplicity $d=2$, and Fourier transform gives
\begin{equation}
n(r)\sim \int dk k n(k)J_0(kr),
\end{equation}
with $J_0$ the Bessel function. Contribution from the singularity of $n(k)$ can be obtained by partial integration. When $n(k)$ has a jump at $k_F$, one has 
\begin{equation}
n(r)\sim \left[\frac{kJ_1(kr)}{r}n(k)\right]_{k_F-\delta}^{k_F+\delta}\sim \frac{\cos( k_Fr-3\pi/4)}{(k_Fr)^{3/2}}.
\end{equation}
When $n(k)$ has a higher order singularity, further partial integration leads to
\begin{equation}
n(r)\sim \left[\frac{kJ_2(kr)}{r^2}\frac{d n(k)}{dk}\right]_{k_F-\delta}^{k_F+\delta}\sim \frac{\cos(k_Fr-5\pi/4)}{(k_Fr)^{5/2}},
\end{equation}
which decays faster than the Fermi liquid result.

\bibliographystyle{apsrev}
\bibliography{strings,refs}

\end{document}